\documentclass[journal]{IEEEtran}

\ifCLASSINFOpdf
\usepackage[pdftex]{graphicx}
\graphicspath{{../pdf/}{../jpeg/}}
\DeclareGraphicsExtensions{.pdf,.jpeg,.png}
\else
\usepackage[dvips]{graphicx}
\graphicspath{{../eps/}}
\DeclareGraphicsExtensions{.eps}
\fi

\usepackage{setspace}
\usepackage{ifpdf}
\usepackage[cmex10]{amsmath}
\usepackage{algorithmic}
\usepackage{array}
\usepackage{mdwmath}
\usepackage{mdwtab}
\usepackage{eqparbox}
\usepackage{amssymb}
\usepackage[bold,full]{complexity}
\usepackage{color}
\usepackage{soul}

\begin{document}

\title{Optimal Transaction Queue Waiting in Blockchain Mining}
\author{\authorblockN{Gholamreza Ramezan$^1$, Cyril Leung$^1$ and Chunyan Miao$^2$\\
$^1$Department of Electrical and Computer Engineering\\
The University of British Columbia, Vancouver, Canada\\
$^2$School of Computer Science and Engineering\\Nanyang Technological University, Singapore\\
Email:\{gramezan, cleung\}@ece.ubc.ca}, ascymiao@ntu.edu.sg\vspace{-1cm}}

\markboth{}
{Shell \MakeLowercase{\textit{et al.}}: Bare Demo of IEEEtran.cls for IEEE Journals}

\maketitle
\begin{abstract}
Blockchain systems are being used in a wide range of application domains. They can support trusted transactions in time critical applications. In this paper, we study how miners should pick up transactions from a transaction pool so as to minimize the average waiting time per transaction. We derive an expression for the average transaction waiting time of the proposed mining scheme and determine the optimum decision rule. Numerical results show that the average waiting time per transaction can be reduced by about $10\%$ compared to the traditional no-wait scheme in which miners immediately start the next mining round using all transactions waiting in the pool. 
\end{abstract}
\IEEEpeerreviewmaketitle
\section{Introduction}\label{sec_intro}\vspace{-.1cm}
\IEEEPARstart{B}{lockchain} systems have been widely deployed as a means to enable asset exchanges among parties who may not trust each other \cite{ramezan2020_tii, ramezan2018_hindawi, Bitcoin}. In a blockchain system, user transactions arrive at a transaction pool to await processing by miners. After a miner completes its current mining round, it picks up all $L$ transactions which arrived at the pool since its last visit and starts generating the next block. We refer to the time interval between the arrival of a transaction and the start of mining of a block in which it is contained as its \emph{waiting time}, denoted by $W$. In this paper, we are interested in minimizing the average waiting time per transaction, $\overline{W}$. If $L$ is relatively small, it may be advantageous for the miner to wait for some more transactions to arrive before starting the mining process. Thus, an interesting question is \emph{how many transactions the miner should wait for.} 

\par Several papers have discussed blockchain mining time. In \cite{BlockchainQueue1}, the average waiting time for transactions is analyzed assuming an exponential distribution for  the block generation process. Queueing models for the Bitcoin \cite{Bitcoin} and Ethereum \cite{Ethereum} blockchains have been proposed in \cite{BlockchainQueue2} and \cite{BlockchainQueue3}. These works assume exponential distribution for the block generation process. In \cite{BlockchainQueue3}, it is  assumed that multiple miners collaborate in generating new blocks. 
\par In contrast to the existing works, we propose a new mining strategy to minimize the average waiting time per transaction. In our proposed scheme, the miner waits for a certain minimum number, $D$, of transactions to be in the pool before starting a new mining round. For convenience, we refer to the scheme as \emph{Wait-Min(D)} and note that the traditional no-wait scheme is simply \emph{Wait-Min}(1). Our contributions are as follows:
\begin{itemize}
\item We propose the \emph{Wait-Min(D)} scheme and derive the average waiting time per transaction.
\item We determine the \emph{optimal} value, $D^*$, which minimizes the average waiting time per transaction. It is found that $D^* \approx 0.9 \lambda / \mu$ where $\lambda$ is the average transaction arrival rate and $\mu$ is the average mining round service rate. 
\item We validate our analysis through simulations. Results show that the average per transaction waiting time for \emph{Wait-Min($D^*$)} is about 10\% lower than \emph{Wait-Min}(1). 
\end{itemize}
\par The remainder of this paper is organized as follows. A queueing model for a blockchain system and the proposed \emph{Wait-Min(D)} scheme are described in Section \ref{sec_sys_model}. In Section \ref{sec_analysis}, we derive the average waiting time per transaction for the \emph{Wait-Min(D)}  scheme. Numerical results are provided in Section~\ref{sec_numerical}. The main findings are summarized in Section~\ref{sec_concl}.
\section{System Model and Proposed Scheme }\label{sec_sys_model}
\subsection{System Model}\label{subsec_sys_model}
\begin{figure}
\centering
\includegraphics[scale=0.45]{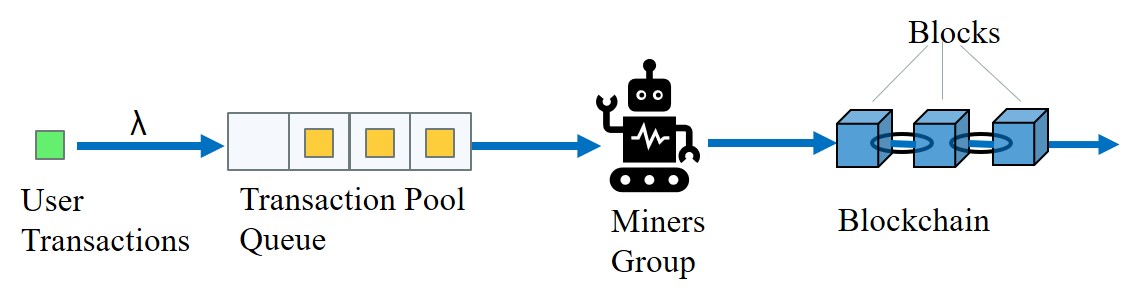}
\caption{Queueing model for a blockchain system.}
\label{fig_system2}
\end{figure}
We consider the queueing model for a blockchain system shown in Fig. \ref{fig_system2}. User transactions arrive at the pool queue according to a Poisson process with arrival rate $\lambda$ and wait there until the miner (server) completes its current mining round. The miner then picks up transactions from the pool and starts generating the next block. The round mining time follows an exponential distribution with average rate $\mu$  \cite{BlockchainQueue1}. Note that in contrast to a traditional queueing system \cite{kleinrock, datanetworks} in which a server serves one customer at a time, the service (mining) time in the blockchain system is \emph{independent} of the number of customers (transactions) in the block. 
\subsection{Proposed Scheme}\label{subsec_holdoff_scheme}
In a traditional blockchain, after a miner completes a mining round, it will pick up all the transactions which are waiting in the pool queue and start a new mining round. In the proposed \emph{Wait-Min(D)} scheme, the miner waits until a minimum of $D \geqslant 1$ transactions are in the queue before it starts a new mining round.
\section{Derivation of the Average Waiting Time}\label{sec_analysis}
\begin{figure*}
\centering
\includegraphics[scale=0.6]{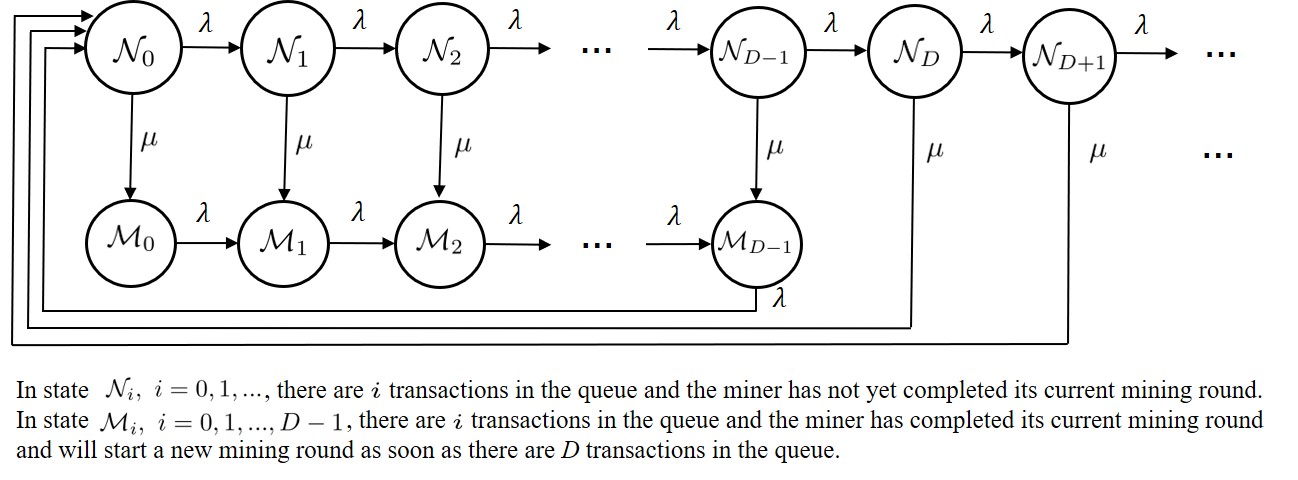}
\caption{State transition diagram for the \emph{Wait-Min(D)} scheme.}
\label{fig_statetransitiondiagram}
\end{figure*}
In this section, we derive analytic expressions for the average waiting time per transaction, $\overline{W}$, for the \emph{Wait-Min(D)} scheme. Fig. \ref{fig_statetransitiondiagram} shows the state transition diagram for the \emph{Wait-Min(D)} scheme. In state $\mathcal{N}_i$, there are $i=0,\ldots,\infty$ transactions in the queue and the miner has not yet completed its current mining round. In state $\mathcal{M}_i$ there are $i=0,..,D-1$ transactions in the queue and the miner (having completed its current mining round) is waiting for a minimum of $D$ transactions in the queue in order to start a new mining round. 

Let $\Pi_{\mathcal{N}_i}$ and $\Pi_{\mathcal{M}_j}$ denote the stationary probabilities of states $\mathcal{N}_i$ and $\mathcal{M}_j$, respectively. Note that the probability, $\Pi_l$, of $l$ transactions in the queue is given by 
\begin{equation}\label{EqS1}
\begin{aligned}
\Pi_{l} = \Pi_{\mathcal{N}_{l}} + \Pi_{\mathcal{M}_{l}}
\end{aligned}
 \end{equation}
We now derive expressions for $\Pi_{\mathcal{N}_{l}}$ and $\Pi_{\mathcal{M}_{l}}$. Equating the flows in and out of state $\mathcal{N}_{i}$ yields 
\begin{equation}\label{EqS2}
\begin{aligned}
(\lambda + \mu) \Pi_{\mathcal{N}_{i}} = \lambda \Pi_{\mathcal{N}_{i-1}},  \ \ i=1,2,3,...
\end{aligned}
 \end{equation}
It follows from (\ref{EqS2}) that 
\begin{equation}\label{EqS3}
\begin{aligned}
\Pi_{\mathcal{N}_{i}}= ( \frac{\lambda}{\lambda+\mu})^i \Pi_{\mathcal{N}_{0}}
\end{aligned}
 \end{equation}
Similarly, equating the flows in and out of state $\mathcal{M}_{i}$ yields
\begin{equation}\label{EqS4}
\begin{aligned}
\lambda \Pi_{\mathcal{M}_{0}} =\mu \Pi_{\mathcal{N}_{0}}
\end{aligned}
 \end{equation}
\begin{equation}\label{EqS5}
\begin{aligned}
\lambda ( \Pi_{\mathcal{M}_{i}} - \Pi_{\mathcal{M}_{i-1}} ) = \mu \Pi_{\mathcal{N}_{i}}, \ \  i=1,2,...,D-1
\end{aligned}
\end{equation}
Using (\ref{EqS4}) and (\ref{EqS5}), we can obtain 
\begin{equation}\label{EqS6}
\begin{aligned}
\Pi_{\mathcal{M}_{i}} = \big ( \frac{\lambda+\mu}{\lambda}- ( \frac{\lambda}{\lambda+\mu} )^i \big )  \Pi_{\mathcal{N}_{0}}, \ \  i=0,1,2,...,D-1
\end{aligned}
 \end{equation}
Using the fact that 
\begin{equation}\label{EqS7}
\begin{aligned}
\sum_{i=0}^{\infty} \Pi_{\mathcal{N}_i} +\sum_{i=0}^{D-1} \Pi_{\mathcal{M}_i} =1
\end{aligned}
 \end{equation}
together with (\ref{EqS3}) and (\ref{EqS6}), we can obtain
 \begin{equation}\label{EqS8}
\begin{aligned}
\Pi_{\mathcal{N}_{0}}=\frac{1}{D(\frac{\lambda+\mu}{\lambda})+\frac{\lambda+\mu}{\mu} (\frac{\lambda}{\lambda+\mu})^{D} }.
\end{aligned}
 \end{equation} 
Let $L$ denote the transaction queue length. The average queue length can be obtained as 
 \begin{equation}\label{EqS9}
\begin{aligned}
\overline{L}=\sum_{l=0}^{\infty} l \ \Pi_l
\end{aligned}
 \end{equation} 
From (\ref{EqS1}), (\ref{EqS3}), and (\ref{EqS6}), $\Pi_l$ can be reduced to 
 \begin{equation}\label{EqS10}
\begin{aligned}
\Pi_l=
\left\{
    \begin{array}{ll}
        \frac{\lambda+\mu}{\lambda} \Pi_{\mathcal{N}_0}, & 0 \leqslant l \leqslant D-1 \\
        (\frac{\lambda}{\lambda+\mu})^l \Pi_{\mathcal{N}_0}, &  l \geqslant D
    \end{array}
\right.
\end{aligned}
 \end{equation} 
Using (\ref{EqS10}) in (\ref{EqS9}) and simplifying, we can write
\begin{equation}\label{Eq13}
\begin{aligned}
\overline{L}= \ & \Pi_{\mathcal{N}_{0}} \bigg ( \big ( \frac{\lambda+ \mu}{\lambda}\big ) \big (\frac{D(D-1)}{2} \big ) \\
&+ \frac{\lambda}{\mu}  \big (\frac{\lambda}{\mu} + D\big ) \big (\frac{\lambda}{\lambda+\mu }\big )^{D-1}    \bigg )
\end{aligned}
 \end{equation} 
We will denote the average queue length by  $\overline{L}(D)$ to explicitly show the dependency on $D$. Using Little's law \cite{kleinrock}, the average waiting time per transaction can be obtained as
\begin{equation}\label{EqAWW}
\begin{aligned}
\overline{W}(D)= \frac{\overline{L}(D)}{\lambda} .
\end{aligned}
\end{equation}
\section{Numerical Results}\label{sec_numerical}
In this section, numerical results based on the expressions derived in Section \ref{sec_analysis} are presented. Computer simulations were used to verify the analytic results. 
\begin{figure}
\centering
\includegraphics[scale=0.55]{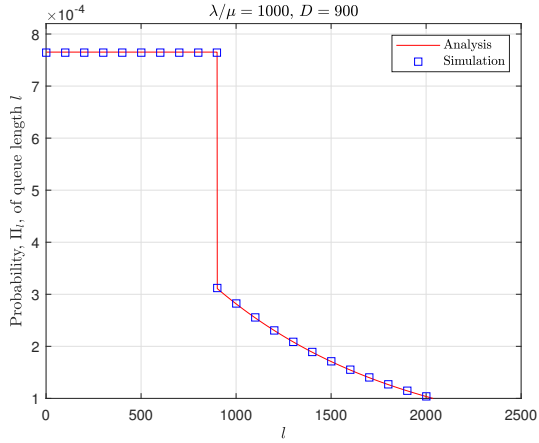}
\caption{Probability of \textit{l} transactions in the queue with \emph{Wait-Min}($D$) for $\lambda / \mu=1000$ and $D = 900$. }
\label{fig_results1}
\end{figure}
\begin{figure}
\centering
\includegraphics[scale=0.55]{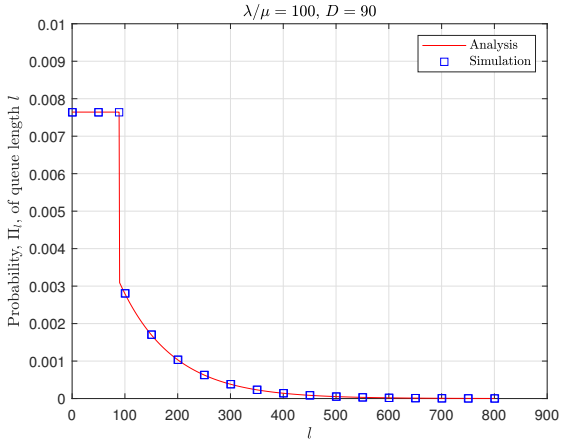}
\caption{Probability of \textit{l} transactions in the queue with \emph{Wait-Min(D)} for $\lambda / \mu=100$ and $D = 90$. }
\label{fig_results2}
\end{figure}
Figs. \ref{fig_results1} and \ref{fig_results2} show the probability, $\Pi_l$, of queue length $l$, obtained using (\ref{EqS1}), (\ref{EqS3}), (\ref{EqS6}), and (\ref{EqS8}) for $\lambda / \mu=1000, \ D = 900$ and $\lambda / \mu=100, \ D = 90$, respectively. It can be observed that $\Pi_l$ drops sharply at $l=D$. This is not surprising since $\lambda \overline{S} > D$ and $D$ is the minimum transactions needed before the miner starts a new mining round.
 
\begin{figure}
\centering
\includegraphics[scale=0.55]{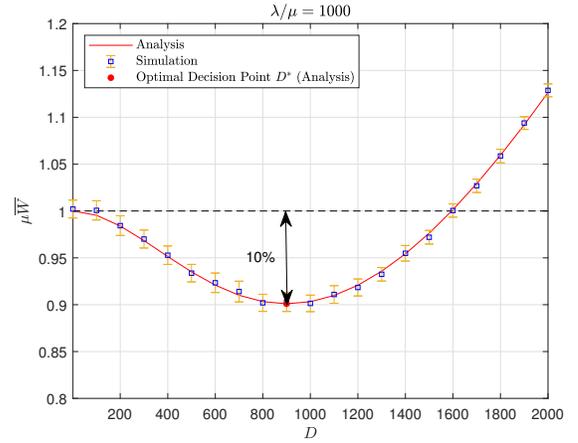}
\caption{Normalized average waiting time per transaction, $\mu \overline{W}$, for \emph{Wait-Min(D)} as a function of $D$ with $\lambda / \mu=1000$.}
\label{fig_results3}
\end{figure}

\begin{figure}
\centering
\includegraphics[scale=0.55]{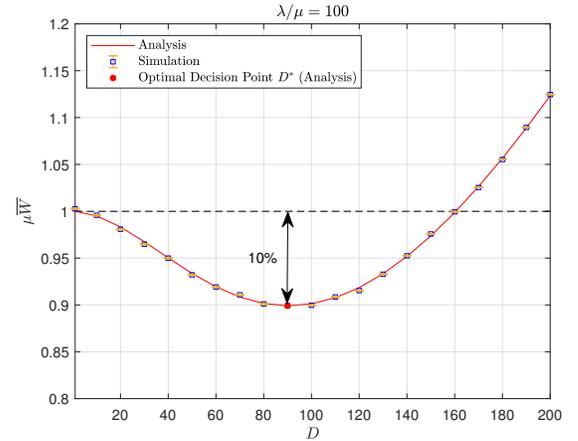}
\caption{Normalized average waiting time per transaction, $\mu \overline{W}$, for \emph{Wait-Min(D)} as a function of $D$ with $\lambda / \mu=100$.}
\label{fig_results4}
\end{figure}

Figs. \ref{fig_results3} and \ref{fig_results4} show the average waiting time per transaction, $\overline{W}$, for $\lambda / \mu=1000$ and $100$, respectively. The optimal values of $D$ which minimize $\overline{W}$ are $D^*=900$ and $90$, respectively. In both cases, the average waiting time per transaction for \emph{Wait-Min($D^*$)} is about 10\% lower than that for the traditional \emph{Wait-Min}(1) scheme.

\begin{figure}
\centering
\includegraphics[scale=0.55]{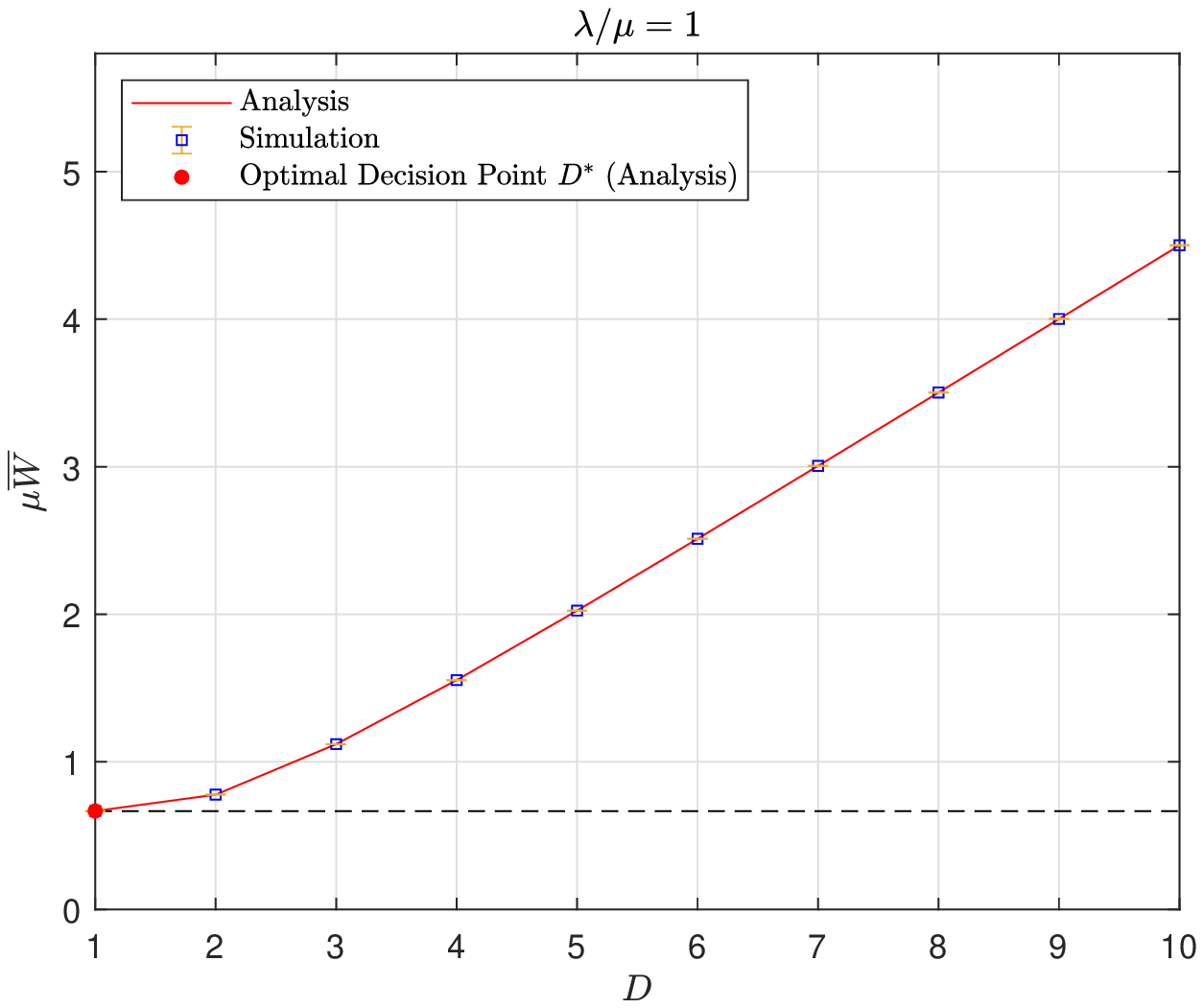}
\caption{Normalized average waiting time per transaction, $\mu \overline{W}$, for \emph{Wait-Min(D)} as a function of $D$ with $\lambda / \mu=1$.}
\label{fig_results6}
\end{figure}

\begin{figure}
\centering
\includegraphics[scale=0.55]{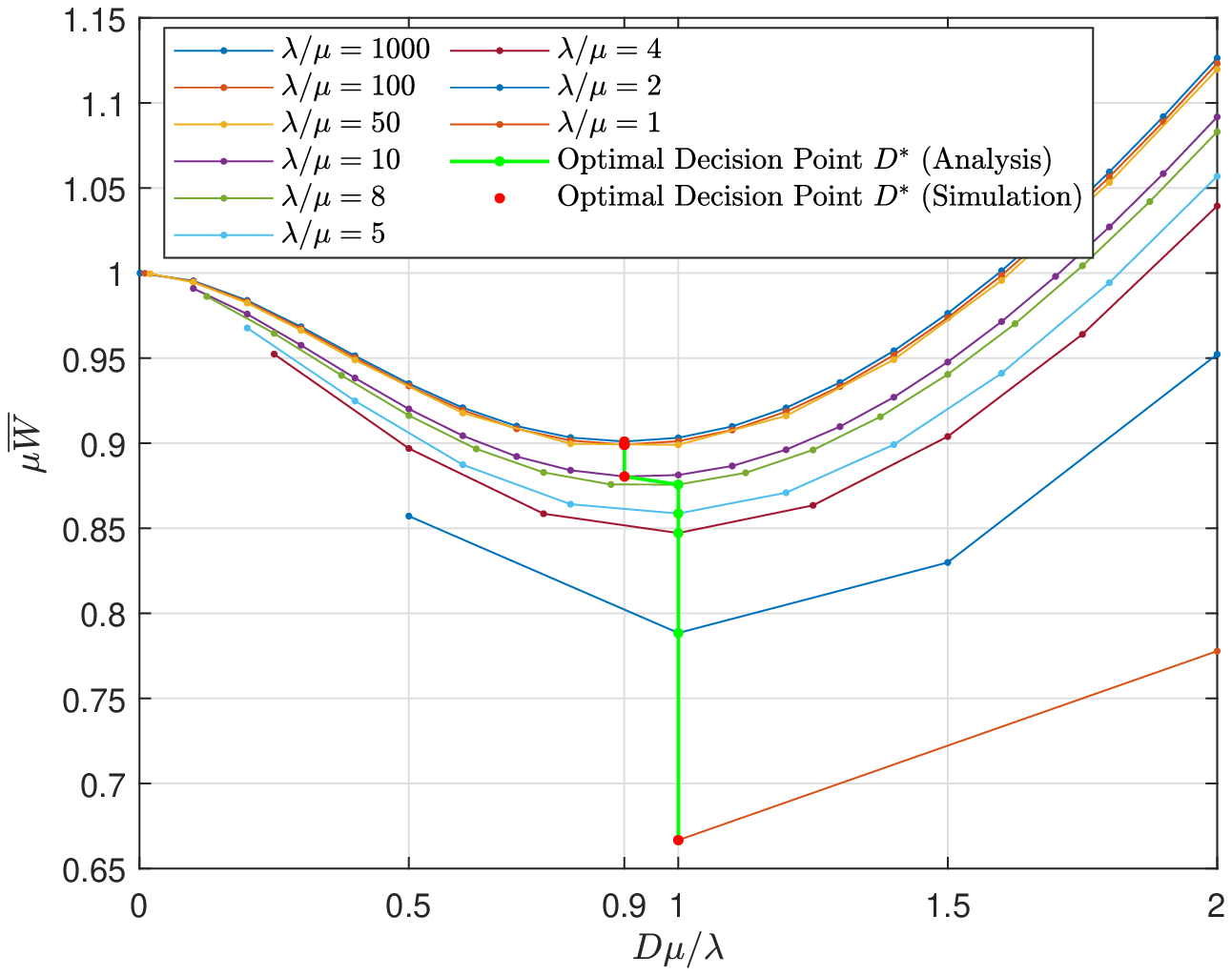}
\caption{Normalized average waiting time $\mu \overline{W}$ as a function of $D \mu / \lambda$. }
\label{fig_results8}
\end{figure}

Fig. \ref{fig_results6} shows $\overline{W}$ as a function of $D$ for $\lambda / \mu=1$. In this case, the traditional \emph{Wait-Min}(1) scheme is optimal. The results are summarized in Fig. \ref{fig_results8} which shows a plot of $\mu \overline{W}$ as a function of $D \mu / \lambda$. For any value of $\lambda/\mu >0$, it was found that the optimal value of $D$ is well approximated by $D^* \approx \lceil 0.9 \lambda / \mu \rceil$ and \emph{Wait-Min($D^*$)} has a $\overline{W}$ value which is about 10\% lower than that of the traditional \emph{Wait-Min}(1) scheme.

\begin{figure}
\centering
\includegraphics[scale=0.55]{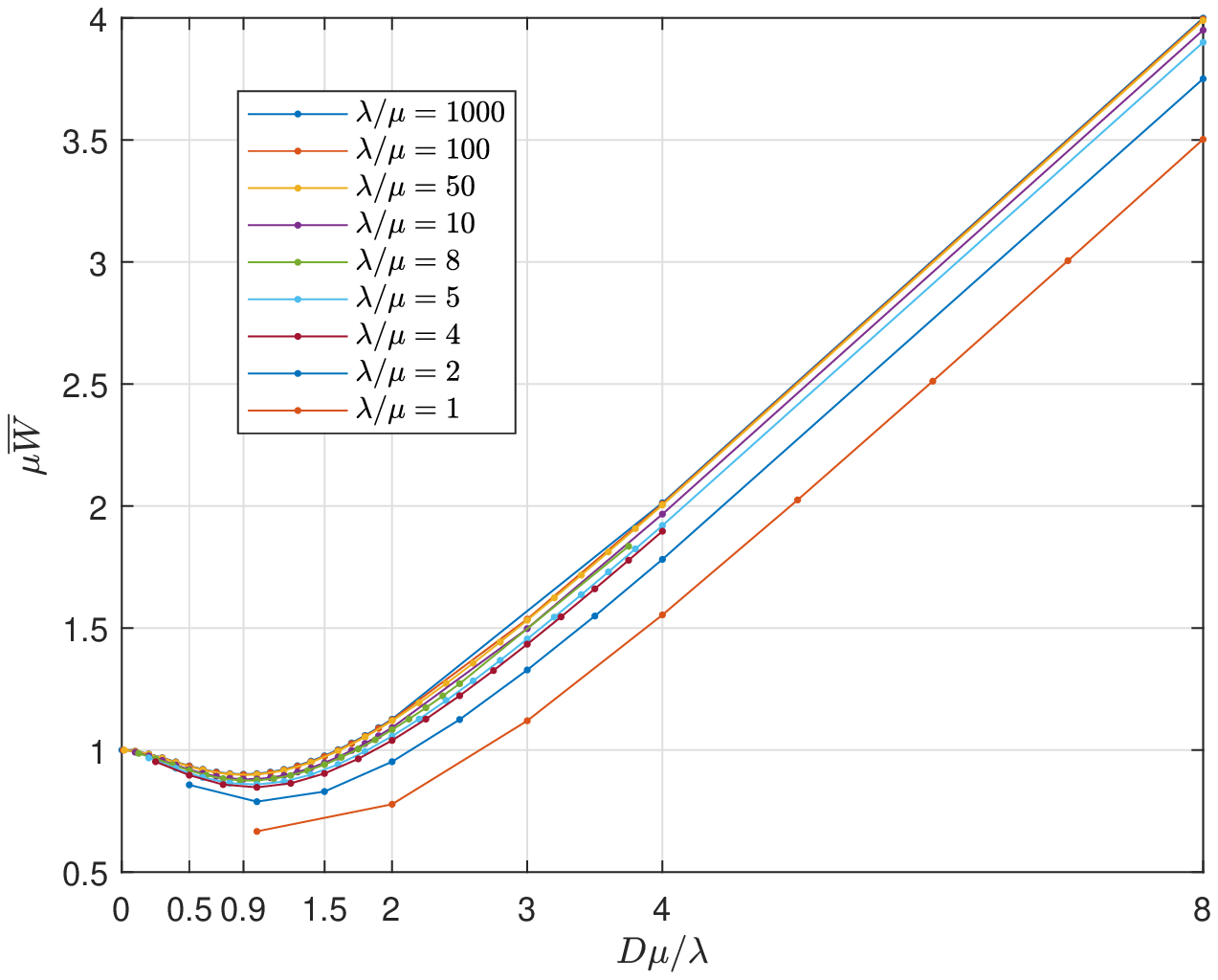}
\caption{Normalized average waiting time $\mu \overline{W}$ as a function of $D \mu / \lambda$.}
\label{fig_results101}
\end{figure}

For large values of $D$, $\overline{W} \approx D/(2\lambda)$ since the first transaction to the queue has an average waiting time of $D/\lambda$ whereas the $D^{th}$ transaction has zero waiting time. This can be observed in Fig. \ref{fig_results101} in which the slopes of all the curves have an asymptotic value of $1/2$.

\begin{figure}
\centering
\includegraphics[scale=0.55]{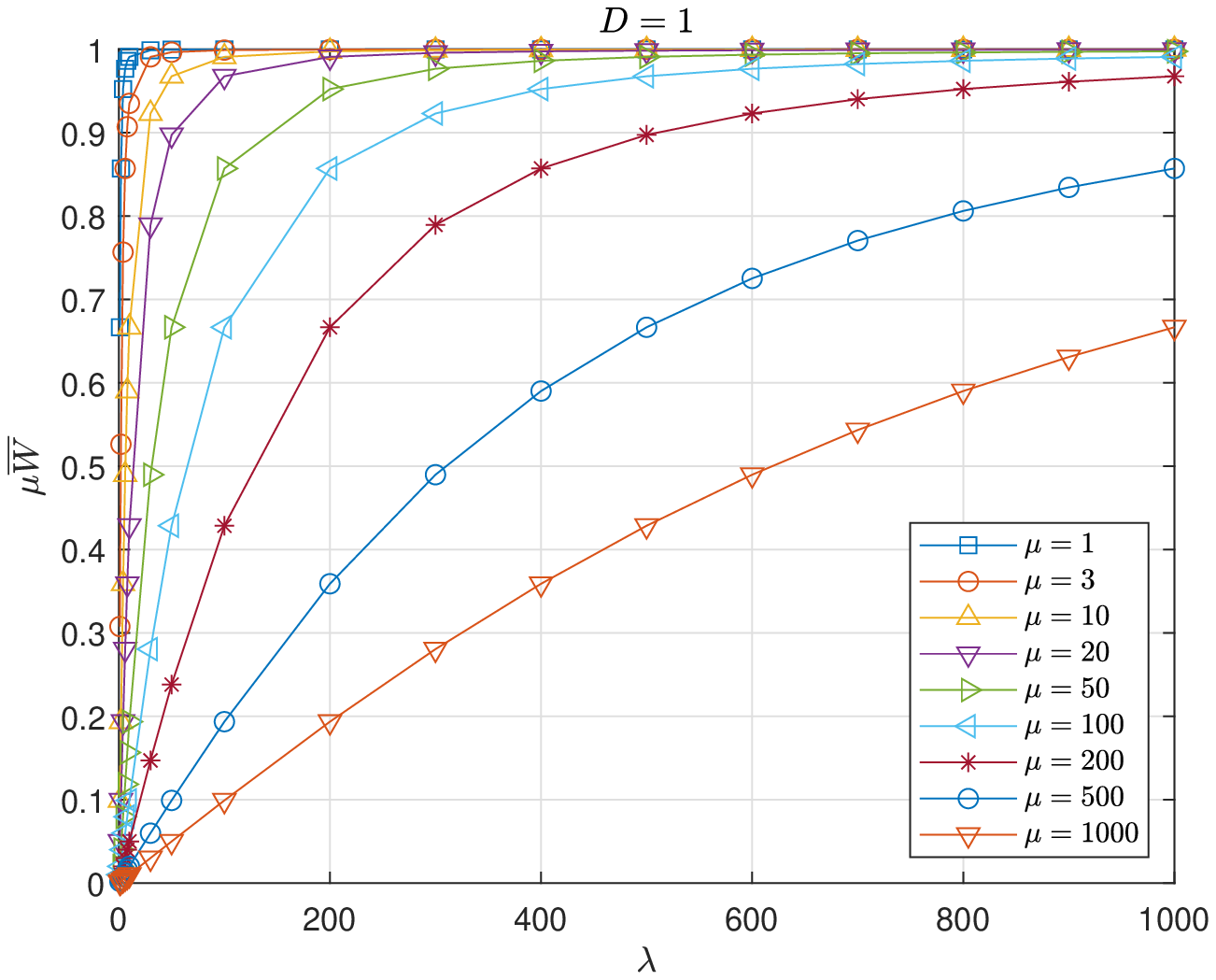}
\caption{Normalized average waiting time $\mu \overline{W}$ as a function of $\lambda$ when $D=1$.}
\label{fig_results102}
\end{figure}

For $D$=1, and using (8), (11), (12), we can write the average waiting time per transaction as 
\begin{equation}\label{examp1}
\begin{aligned}
\overline{W}(1)=\frac{\lambda(\lambda+\mu)}{\mu(\lambda\mu+\mu^2+\lambda^2)}.
\end{aligned}
\end{equation}
Fig. \ref{fig_results102} shows the normalized average waiting time per transaction for $D=1$ (i.e., \emph{Wait-Min}(1)) for different values of $\lambda$ and $\mu$. As to be expected, for a fixed value of $\mu$, $\mu \overline{W}$ increases with $\lambda$ and approaches 1 for $\lambda \gg \mu$.
\section{Conclusion}\label{sec_concl}\vspace{-.1cm}
The \emph{Wait-Min(D)} scheme was proposed to minimize the average waiting time per transaction, $\overline{W}$, in a blockchain system. An analytic expression for $\overline{W}$ was derived and verified using simulations. Numerical results show that the proposed scheme can reduce $\overline{W}$ by about 10\% compared with the traditional scheme in practical settings.
\ifCLASSOPTIONcaptionsoff
  \newpage
\fi
\bibliographystyle{IEEEtran}
\bibliography{biblio}
\end{document}